\documentclass[11pt]{article}
\usepackage{moriond,epsfig}

\def\Journal#1#2#3#4{{#1} {\bf #2}, #3 (#4)}

\def\AA{\em A\&A}
\def\ApJ{\em ApJ}
\def\ApJS{\em ApJS}

\def\be{\begin{equation}}
\def\ee{\end{equation}}
\def\bea{\begin{eqnarray}}
\def\eea{\end{eqnarray}}

\begin{document}
\vspace*{4cm}
\title{SOFT GAMMA-RAY BURSTS IN THE BATSE DATA}
\author{YA.TIKHOMIROVA${}^1$, B.E.STERN${}^{2,1}$,\\
A.KOZYREVA${}^3$, J.POUTANEN${}^4$ }

\address{${}^{1}$ Astro Space Centre of
Lebedev Physical Institute,\\
Profsoyuznaya 84/32, Moscow 117997, Russia;\\
${}^{2}$ Institute for Nuclear Research, Moscow, Russia;\\
${}^{3}$ Sternberg Astronomical Institute, Moscow, Russia;\\
${}^{4}$ Astronomy Division, FIN-90014 University of Oulu, Finland}

\maketitle\abstracts{
We performed the scan of the BATSE DISCLA records
inspecting 25-50 keV range to pick up soft GRBs.
We applied the same technique as in our previous
(Stern et al. 2001)\cite{stern} scan in the 50-300 keV range.
We scanned about 1.8 year of the data
and found 30 new gamma-ray burst (GRB)s.
The total number of soft GRBs in the BATSE data,
with the count rate in the 25-50 keV range higher
than in the 50-300 keV range, is only 4\%.
There are about two extremely soft events per year
which are invisible about 50 keV and which could constitute a
new class of events separate from the GRB phenomenon.
We also show that X-ray flashes (XRFs)
detected by BeppoSAX have a hardness ratio
very similar to that of normal GRBs, supporting a view
that  XRFs and GRBs are a single phenomenon
with a wide spectral variety.
 }

\section{Introduction}

Observations of the prompt gamma-ray bursts (GRBs) emission
by different instruments show that
their spectra can extend from several keV up to a few MeV,
sometimes up to GeV range.
Particularly, BeppoSAX have detected X-ray
dominated bursts which were named
X-ray flashes (XRFs)\cite{heise}.
To clarify the complete picture
of GRBs spectral variety, simultaneous broadband observations
with different instruments are required. By now the data
with such kind of observations
\cite{heise,kippen,hetebar,hetesak} are rather poor
and their interpretation is complicated because of
different instrument responses, associated biases and
insufficient statistics at some spectral ranges.

The records of Burst And Transient Sources Experiment
(BATSE)\cite{batse}   all-sky 9.1 year
continuous monitoring in gamma-range
give unique possibility for combined analysis
with X-ray observations of GRB prompt emission. In the
BATSE most sensitive gamma-detectors over the GRB history
were used. Only recently lunched SWIFT
experiment\cite{swift} has a more sensitive gamma-detector.
However, during next several years SWIFT
can not accumulate so much statistics as BATSE.

The BATSE detectors were sensitive to photons
from about 25 keV up to about 1 MeV.
However, the on-board trigger identified GRBs according
to the signal  in the 50-300 keV range only.
GRBs with a soft spectrum can be lost
because the 25-50 keV range was not inspected.
Such events can be found in  BATSE records.
We use BATSE data to search for soft
GRBs and analysed the BATSE results together with the
recent observations of soft GRBs and  XRFs
by other instruments.

\section{Search for soft GRBs in the BATSE data}

We have performed the scan of the BATSE data
with trigger on 25-100 keV  to pick up soft GRBs.
The continuous daily 1.024 s time
resolution DISCLA records of count rate
in 8 BATSE detectors in 4 energy channels
(25-50, 50-100, 100-300 and 300-1000 keV) were used.
We applied the same technique and the same algorithm
as in our previous scan of the BATSE DISCLA records
in the 50-300 keV range\cite{stern}.
The number of active, variable sources is much larger
in the 25-50 keV range than in the
harder energy range.
So we paid additional attention of identification of
GRB-like events.

\begin{figure}
\hskip 0.01cm
\psfig{figure=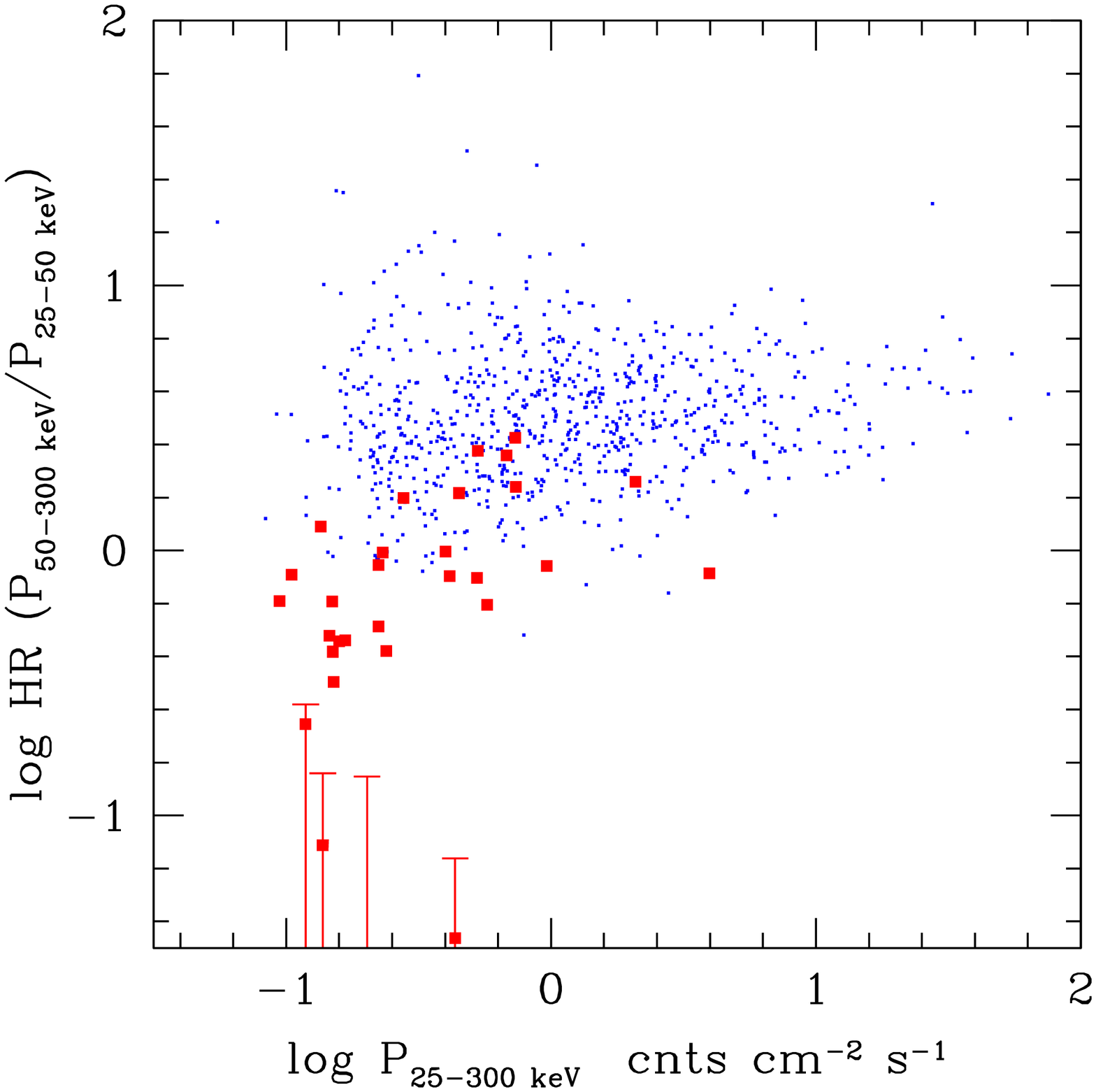,height=7.7cm}
\hskip 0.5cm
\psfig{figure=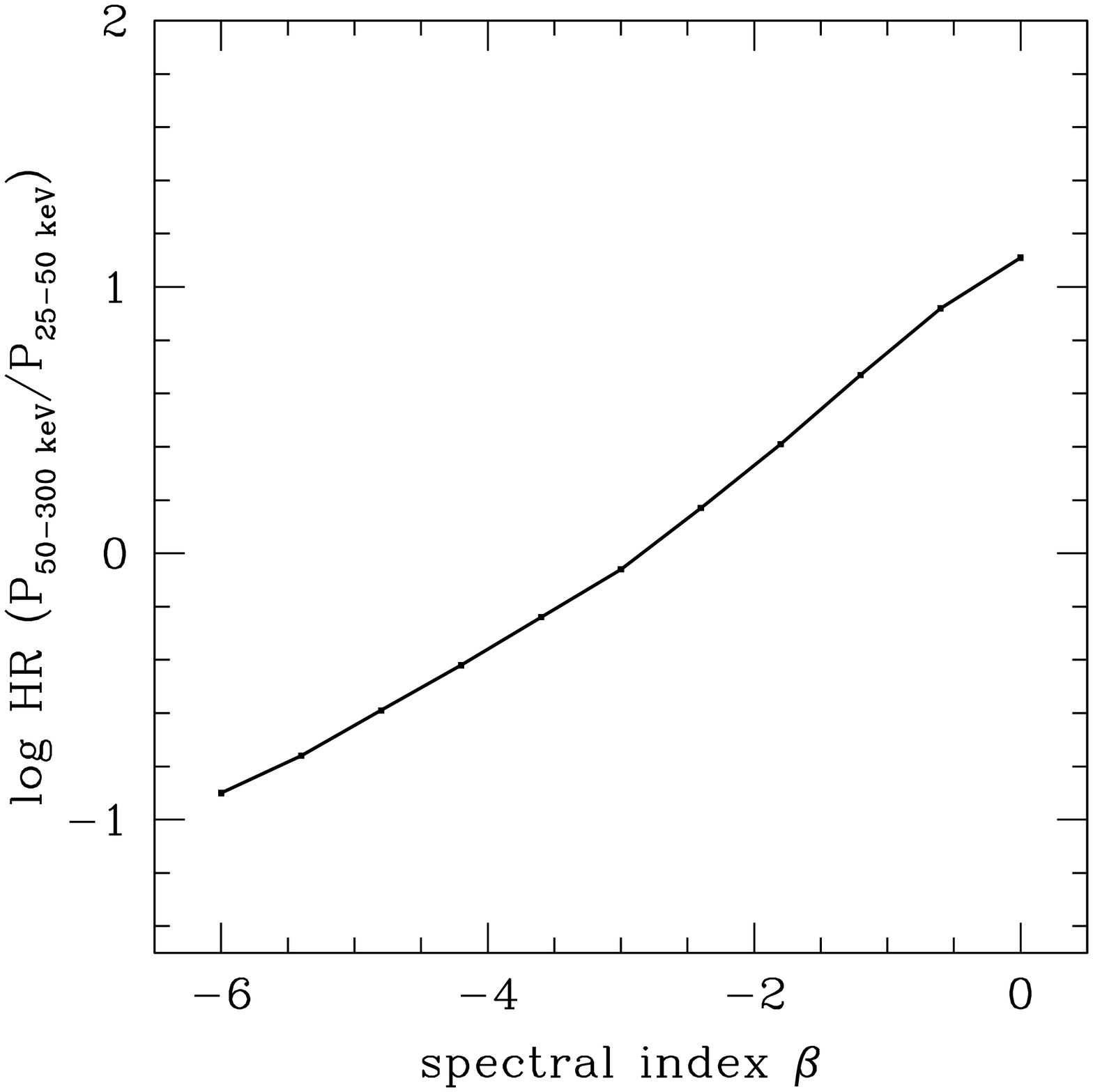,height=7.7cm}
\caption{(left) Hardness-intensity diagram of GRBs
found in the BATSE data for the time period
TJD 11000-11699. Dots: the scan
in the 50-300 keV range\protect\cite{stern},
squares: the present scan in the 25-100 keV range.}
\label{fig:ha}
\end{figure}

\begin{figure}
\caption{(right) Hardness ratio of simulated GRBs with
the power-law spectrum (accounting for the BATSE detector response matrix$^8$)
versus the photon spectral index $\beta$.
\label{fig:habeta}}
\end{figure}

We scanned the data of about 1.8 year
(TJDs 11000-11699) and found 30 new GRB-like events.
For the same time period there are about 800 GRBs
found by our previous scan in 50-300 keV range\cite{stern}.
Hardness-intensity diagram (Fig.~1)
shows that although GRBs of a new sample
are softer on average, these samples do overlap.
There are 22 GRBs in a new sample and
15 GRBs in the old sample\cite{stern} with the
count rate in the 25-50 keV band
higher than that in the 50-300 keV band.
We consider these 37 events as a sample
of soft BATSE GRBs. If we approximate
their spectra with a power-law (assuming $E_{\rm peak}$
below 25 keV) then this criterion of softness
corresponds to the photon power-law index $\beta<-3$
(see Fig.~2). Soft GRBs
are isotropically distributed on the sky
and have typical for GRBs durations and
light curves (Fig.~3).

\begin{figure}
\hskip 2.5cm
\psfig{figure=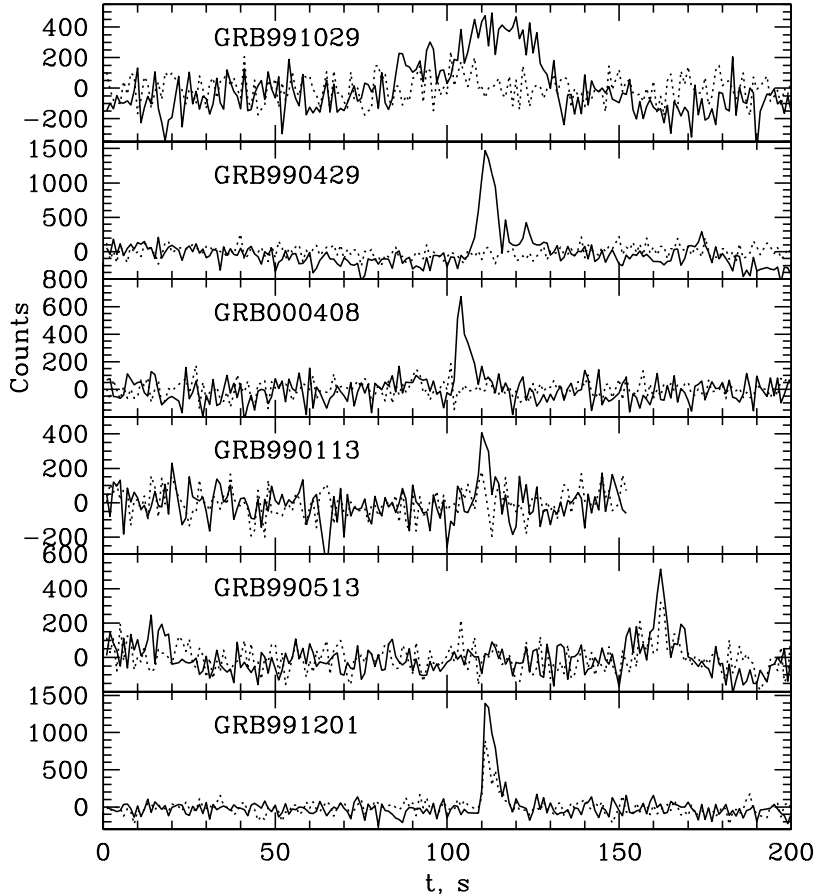,height=12cm}
\caption{Fragments of the BATSE records with the softest GRBs
found in the 25-100 keV range. Solid curve shows the 25-50 keV, while
the dashed curve the 50-100 keV count rates.
\label{fig:prof}}
\end{figure}

\section{Spectral properties of XRFs/GRBs using BATSE/BeppoSAX/HETE2 data}

Our scan (as well as an alternative scan\cite{kom}) in
the BATSE records in the 25-50 keV range  has yielded
surprisingly small number of new soft GRBs.
 One could expect to see XRFs tails in the
BATSE range as soft bursts.
However, according to Figs.~1 and 2,
an event gives a larger signal above 50 keV,
if XRFs spectra have the high energy
power-law index $\beta > -3$.

Probably this is the case for most of XRF.
Recently, BeppoSAX XRFs were identified
in the BATSE data\cite{kippen} (most of them
were detected earlier by Stern et al. as GRBs\cite{stern}).
These events, indeed, have a high hardness ratio
(see Fig.~4).

\begin{figure}
\hskip 3.5cm
\psfig{figure=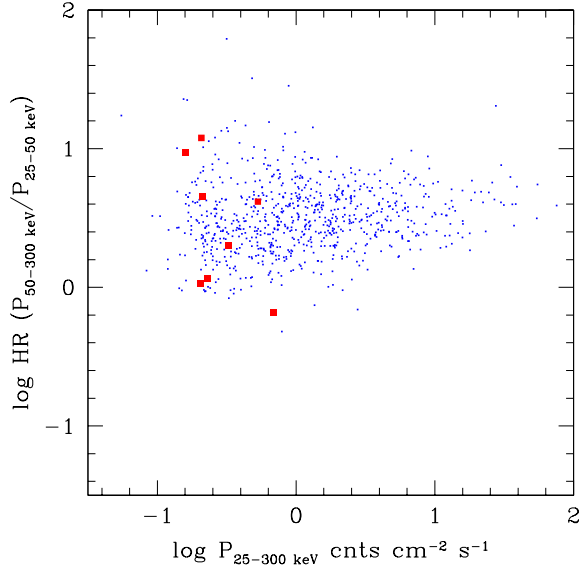,height=7.7cm}
\caption{Hardness-intensity diagram of the BATSE counterparts of XRFs
detected by BeppoSAX (squares) in comparison with GRBs
found by our scan in the 50-300 keV
range\protect\cite{stern}, TJD 11000-11699 (dots).}
\label{fig:haSAX}
\end{figure}

We observe in BATSE data a small number
(2 per year with flux down to 0.1 ph cm$^2$ s$^{-1}$)
of really soft outlying events
with no detectable flux above 50 keV.
They probably have an  exponential cutoff in the
spectra below $\sim 15$ keV. Correspondingly, in HETE-2
data\cite{hetesak} there is one outlying event
with exponential cutoff at $\sim 8$ keV (GRB~020625).

\section{Conclusions}

1. XRFs spectra usually have
a comparatively hard ($\beta >-3$) high energy tail.
Their hardness distribution is similar to that of GRBs.
These results support the opinion that XRFs
and GRBs are a single phenomenon
with a wide variety of spectra.
\\
2. The number of soft GRBs in the BATSE data,
with the count rate in the 25-50 keV range higher
than in the 50-300 keV range, is only 4\%
(20 per year with flux down to 0.1 ph cm$^2$ s$^{-1}$).\\
3. There exist X-ray transients (2 per year) with the
spectral   cutoff below $\sim 15$ keV
which could constitute a  class of events separate
from the GRB phenomenon.

\section*{Acknowledgments}
This work is supported by NORDITA
Nordic project
in high energy astrophysics in the INTEGRAL era,
Russian Fondation for Basic Research (grant 04-02-16987),
and Russian Fondation of Science Support (Y.T.).
We thank also the conference organizers for financial support of our
participation.


\section*{References}

\begin{thebibliography}{99}

\bibitem{stern} B.Stern {\it et al}, \Journal{\ApJ}{563}{80}{2001}

\bibitem{kippen} Mc.Kippen {\it et al.}, astro-ph/0102277

\bibitem{heise} J.Heise {\it et al.}, astro-ph/0111246


\bibitem{hetebar} C.Barraud {\it et al.}, \Journal{\AA}{400}{1021}{2003}.

\bibitem{hetesak} T.Sakamoto {\it et al.}, astro-ph/0409128

\bibitem{batse} G.Fishman {\it et al.}, Proc. GRO Science Workshop, 2, 1989

\bibitem{swift} N.Gehrels {\it et al}, \Journal{\ApJ}{611}{1005}{2004}


\bibitem{drm} G.Pendleton {\it et al}, \Journal{\ApJ}{512}{362}{1999}

\bibitem{kom} J.Kommers {\it et al}, \Journal{\ApJS}{134}{385}{2001}
\end{thebibliography}
\end{document}